\documentclass[sigconf]{acmart}

\AtBeginDocument{%
  }

\usepackage{algorithm}
\usepackage{graphicx}
\usepackage{textcomp}
\usepackage{xcolor}
\usepackage{multirow}
\usepackage{makecell}
\usepackage{booktabs}
\usepackage{caption}
\usepackage{subcaption}
\usepackage{tikz}
\usepackage{float}
\usepackage{comment}

\copyrightyear{2026}
\acmYear{2026}
\setcopyright{cc}
\setcctype{by}
\acmConference[GLSVLSI '26]{Great Lakes Symposium on VLSI 2026}{June 22--24, 2026}{Canandaigua, NY, USA}
\acmBooktitle{Great Lakes Symposium on VLSI 2026 (GLSVLSI '26), June 22--24, 2026, Canandaigua, NY, USA}
\acmDOI{10.1145/3787109.3815298}
\acmISBN{979-8-4007-2431-2/2026/06}

\begin{document}

\title{FAPlace: Joint Optimization of Chiplet Placement and Interposer Footprint for 2.5D Systems}

\author{Yubo Hou}
\affiliation{%
  \institution{Institute for Infocomm Research, Agency for Science, Technology and Research (A*STAR)}
  \city{Singapore}
  \country{Singapore}}
\email{Hou_Yubo@a-star.edu.sg}

\author{Sezin Kircali Ata}
\affiliation{%
  \institution{Institute for Infocomm Research, Agency for Science, Technology and Research (A*STAR)}
  \city{Singapore}
  \country{Singapore}}
\email{Ata_Kircali_Sezin@a-star.edu.sg}

\author{Gen Liang Lim}
\affiliation{%
  \institution{Institute for Infocomm Research, Agency for Science, Technology and Research (A*STAR)}
  \city{Singapore}
  \country{Singapore}}
\email{lim_gen_liang@a-star.edu.sg}

\author{Richard Chang}
\affiliation{%
  \institution{Institute for Infocomm Research, Agency for Science, Technology and Research (A*STAR)}
  \city{Singapore}
  \country{Singapore}}
\email{Richard_Chang@a-star.edu.sg}

\author{Mihai Dragos Rotaru}
\affiliation{%
  \institution{Institute of Microelectronics, Agency for Science, Technology and Research (A*STAR)}
  \city{Singapore}
  \country{Singapore}}
\email{Mihai_Dragos_Rotaru@a-star.edu.sg}

\author{Rahul Dutta}
\affiliation{%
  \institution{Institute of Microelectronics, Agency for Science, Technology and Research (A*STAR)}
  \city{Singapore}
  \country{Singapore}}
\email{duttar@a-star.edu.sg}

\author{Ashish James}
\affiliation{%
  \institution{Institute for Infocomm Research, Agency for Science, Technology and Research (A*STAR)}
  \city{Singapore}
  \country{Singapore}}
\email{ashish_james@a-star.edu.sg}

\begin{abstract}
The placement of chiplets on a silicon interposer is a pivotal step in 2.5D system integration, yet existing placement approaches typically assume a pre-defined interposer footprint. This creates a circular dependency: the optimal footprint cannot be known without first solving the placement, while the placement itself is constrained by the given dimensions. An undersized interposer may exclude feasible placements, while an oversized one yields unnecessarily sparse solutions. Moreover, even when the footprint area is minimized, few existing approaches explicitly control the interposer's aspect ratio. To jointly address these challenges, we propose FAPlace, a footprint aware mask guided sequential placement framework. FAPlace operates on a sufficiently large canvas, eliminating the circular dependency by allowing the optimal interposer footprint to emerge as an output of the optimization rather than a pre-specified input. At its core is a novel footprint mask that fuses area compactness with an aspect ratio penalty into a unified spatial cost map. Integrated with wirelength and thermal guidance masks, FAPlace delivers holistic multi-physics optimization in a deterministic, single pass process. Experimental results demonstrate that FAPlace reduces wirelength and footprint area while achieving near-unity aspect ratios, without compromising on thermal performance.
\end{abstract}\vspace{-5pt}

\ccsdesc[500]{Hardware~Physical design (EDA)}
\ccsdesc[300]{Hardware~Electronic design automation}

\keywords{2.5D System, Chiplet Placement, Aspect Ratio, Interposer Footprint, Multi-physics Optimization.}

\maketitle

\vspace{-5pt}
\section{Introduction}
The growing demand for heterogeneous, high-performance computing has propelled 2.5D chiplet integration to the forefront of advanced packaging~\cite{10137172}. By assembling pre-verified chiplets on a silicon interposer, 2.5D systems deliver superior performance and design reusability at a fraction of the monolithic development cost~\cite{sangiovanni2023automated}. However, the physical placement of chiplets on the interposer is a multi-faceted optimization problem that must simultaneously address wirelength and peak temperature~\cite{9643660}.

\begin{figure}[htbp]
  \centering
  \resizebox{0.8\columnwidth}{!}{

\definecolor{cA}{HTML}{6C9BD2}
\definecolor{cB}{HTML}{F4A460}
\definecolor{cC}{HTML}{7BC67E}
\definecolor{cD}{HTML}{B08ED2}
\definecolor{cE}{HTML}{E8D44D}
\definecolor{cF}{HTML}{E07070}
\definecolor{waste}{HTML}{EDEDED}

\begin{tikzpicture}[
    chip/.style={draw=black, line width=0.5pt, rounded corners=0.5pt},
    interposer/.style={draw=black, line width=1.4pt},
    canvas/.style={draw=black!25, line width=0.7pt, dashed},
    lbl/.style={font=\huge\sffamily},
    plbl/.style={font=\huge\sffamily, align=center},
    ttl/.style={font=\huge\sffamily\bfseries, align=center},
    clbl/.style={font=\huge\sffamily, text=black!70},
]


\begin{scope}[shift={(1.4, 0)}]
  \draw[interposer] (0,0) rectangle (3.2,2.4);

  \fill[cA, chip] (0.1,0.1) rectangle (1.5,1.3);
  \node[clbl] at (0.8,0.7) {$c_1$};

  \fill[cF, chip] (1.6,0.1) rectangle (2.4,1.1);
  \node[clbl] at (2.0,0.6) {$c_6$};

  \fill[cD, chip] (1.1,1.4) rectangle (2.1,2.4);
  \node[clbl] at (1.6,1.9) {$c_4$};

  \fill[cE, chip] (2.5,1.1) rectangle (3.2,1.8);
  \node[clbl] at (2.85,1.45) {$c_5$};

  \fill[cC, chip, fill opacity=0.75] (0.1,1.4) rectangle (1.0,2.6);
  \node[clbl] at (0.55,2.0) {$c_3$};
  \draw[red, line width=1.0pt, dashed] (-0.05,2.4) -- (1.05,2.4);
  \draw[red, line width=1.4pt] (0.35,2.55) -- (0.55,2.35);
  \draw[red, line width=1.4pt] (0.35,2.35) -- (0.55,2.55);

  \fill[cB, chip, fill opacity=0.75] (2.5,0.1) rectangle (3.5,0.8);
  \node[clbl] at (3.0,0.45) {$c_2$};
  \draw[red, line width=1.0pt, dashed] (3.2,-0.05) -- (3.2,0.95);
  \draw[red, line width=1.4pt] (3.35,0.55) -- (3.55,0.35);
  \draw[red, line width=1.4pt] (3.35,0.35) -- (3.55,0.55);

  \node[lbl, red!70!black, align=left, anchor=west] at (3.8, 1.2)
    {Feasible placements\\excluded};

  \node[plbl] at (1.6, -0.4) {(a) Existing approaches: pre-defined undersized footprint};
\end{scope}

\begin{scope}[shift={(0, -3.6)}]
  \fill[waste] (0,0) rectangle (6.0,2.4);
  \draw[interposer] (0,0) rectangle (6.0,2.4);

  \fill[cA, chip] (0.2,0.6) rectangle (1.6,1.8);
  \node[clbl] at (0.9,1.2) {$c_1$};

  \fill[cB, chip] (2.2,0.15) rectangle (3.2,0.85);
  \node[clbl] at (2.7,0.5) {$c_2$};

  \fill[cC, chip] (2.0,1.05) rectangle (2.9,2.25);
  \node[clbl] at (2.45,1.65) {$c_3$};

  \fill[cD, chip] (3.5,0.7) rectangle (4.5,1.7);
  \node[clbl] at (4.0,1.2) {$c_4$};

  \fill[cE, chip] (4.7,1.5) rectangle (5.4,2.2);
  \node[clbl] at (5.05,1.85) {$c_5$};

  \fill[cF, chip] (5.0,0.2) rectangle (5.8,1.2);
  \node[clbl] at (5.4,0.7) {$c_6$};

  \node[lbl, red!70!black, align=left, anchor=west] at (6.3, 1.2)
    {Sparse placement\\$+$ extreme AR};

  \node[plbl] at (3.0, -0.4) {(b) Existing approaches: pre-defined oversized footprint};
\end{scope}

\begin{scope}[shift={(0.65, -8.2)}]
  \draw[canvas] (-0.5,-0.5) rectangle (5.2,3.4);
  \node[lbl, black!25, anchor=north east] at (5.1,3.3) {\textit{canvas}};

  \fill[cA, chip] (0.8,0.3) rectangle (2.2,1.5);
  \node[clbl] at (1.5,0.9) {$c_1$};

  \fill[cF, chip] (2.3,0.3) rectangle (3.1,1.3);
  \node[clbl] at (2.7,0.8) {$c_6$};

  \fill[cE, chip] (3.2,0.3) rectangle (3.9,1.0);
  \node[clbl] at (3.55,0.65) {$c_5$};

  \fill[cC, chip] (0.8,1.6) rectangle (1.7,2.8);
  \node[clbl] at (1.25,2.2) {$c_3$};

  \fill[cD, chip] (1.8,1.6) rectangle (2.8,2.6);
  \node[clbl] at (2.3,2.1) {$c_4$};

  \fill[cB, chip] (2.9,1.4) rectangle (3.9,2.1);
  \node[clbl] at (3.4,1.75) {$c_2$};

  \draw[draw=green!55!black, line width=1.3pt]
    (0.7,0.2) rectangle (4.0,2.9);

  \node[lbl, green!45!black, align=left, anchor=west] at (4.2, 1.5)
    {Compact placement\\$+$ near-unity AR};

  \node[plbl] at (2.35, -0.9)
    {(c) FAPlace: sufficiently large canvas, optimized footprint};
\end{scope}

\end{tikzpicture}}\\[-5pt]
  \caption{Motivation of FAPlace. Existing approaches are bottlenecked by pre-defined interposer dimensions, where (a) an undersized footprint limits feasibility and (b) an oversized one leads to sparse placements and extreme aspect ratios (AR). (c) FAPlace paradigm shifts to a sufficiently large canvas, utilizing a footprint mask to output a compact, near-square layout without requiring a predefined layout.} \vspace{-5pt}
  \label{fig:motivation}
\end{figure}
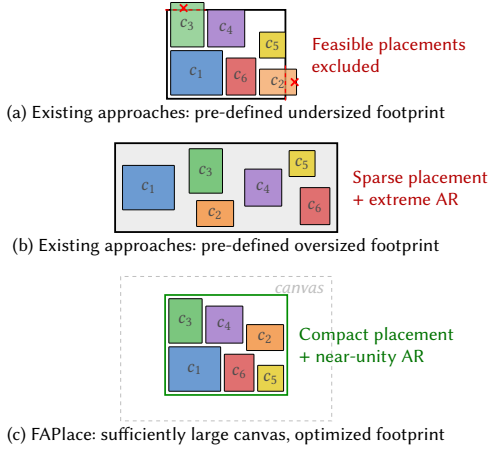


A diverse set of methodologies has been proposed to tackle this challenge, including simulated annealing~\cite{9474011,11311107}, sequence-pair-based methods~\cite{10044783}, analytical formulations~\cite{10.1145/3676536.3676648,yu2025tacplace}, and reinforcement learning~\cite{11392651,10546812}. Despite their diversity in optimization strategy, these methods share a common and fundamental limitation in that they all require a pre-defined interposer footprint as input.

This requirement creates a circular dependency in the design flow. The optimal footprint cannot be known without first solving the placement, while the placement itself is constrained by the given dimensions. As shown in Figure~\ref{fig:motivation}, an undersized interposer may exclude feasible placements, while an oversized one yields unnecessarily sparse solutions. In practice, designers resort to manual trial-and-error or coarse parameter sweeps over canvas sizes, neither of which can efficiently explore the joint space of interposer dimensions and chiplet configurations. We advocate a paradigm shift in which the interposer footprint becomes an output of the placement optimization rather than an input.


Moreover, even when footprint area is minimized on an oversized interposer, an equally important geometric property neglected by all prior works, is the aspect ratio (AR) of the interposer. In practice, industry standard 2.5D packages almost universally adopt near-square interposer form factors~\cite{9501797}. Furthermore, large area interposers are known to suffer from warpage due to CTE mismatch~\cite{6575677}, which an elongated geometry may exacerbate along its longer axis. However, minimizing area alone does not control the layout shape and may yield impractical geometries with extreme aspect ratios.

To jointly address these challenges, we propose FAPlace, a footprint aware mask guided sequential placement framework for 2.5D systems. FAPlace operates on a sufficiently large canvas and employs a novel footprint mask that fuses area compactness with an aspect ratio penalty into a unified spatial cost map. This mask simultaneously drives chiplets toward tight packing and steers the bounding box geometry toward a target shape. Integrated with wirelength and thermal masks, FAPlace delivers holistic multi-physics optimization in a deterministic, single pass procedure.

The specific contributions of this work are:
\begin{itemize}
    \item We introduce FAPlace, which eliminates the circular dependency by allowing the optimal interposer footprint to emerge as an output of the optimization rather than a pre-defined input.
    \item We are the first to incorporate interposer aspect ratio as an explicit optimization objective in 2.5D chiplet placement.
    \item We propose the footprint mask, a unified spatial cost map that jointly minimizes footprint area and penalizes aspect ratio deviation, enabling simultaneous area compactness and shape compliance.
    \item Experiments demonstrate that FAPlace reduces wirelength and footprint area while achieving near-unity aspect ratios, without compromising thermal performance.
\end{itemize}\vspace{-7pt}

\section{Problem Formulation}

Given a set of chiplets $\mathcal{C} = \{c_1, \dots, c_N\}$, where each chiplet $c_i$ is characterized by its width $w_i$, height $h_i$, and thermal design power $p_i$, along with a netlist $\mathcal{N}$ encoding the logical connectivity and a set of pin clumps $\mathcal{K}$, the goal is to determine placement coordinates $\mathcal{X} = \{(x_i, y_i)\}_{i=1}^{N}$ and orientations $\Theta = \{\theta_i\}_{i=1}^{N}$, $\theta_i \in \{0^\circ, 90^\circ\}$, that jointly optimize the following objectives.

\textbf{Wirelength.}
Adopting the flow-based formulation of~\cite{9474011}, we define a variable set $\mathcal{F}$ where $f^n_{iljk}$ quantifies the connections of net $n$ routed from pin clump $l$ on chiplet $i$ to pin clump $k$ on chiplet $j$. The total wirelength is computed as
\begin{equation}
    WL(\mathcal{X}, \mathcal{F}) = \sum_{i,j \in \mathcal{C}} \sum_{l,k \in \mathcal{K}} \sum_{n \in \mathcal{N}} d_{iljk}(\mathcal{X}) \cdot f^n_{iljk},
\end{equation}
where $d_{iljk}$ is the Manhattan distance between the corresponding pin clumps.

\textbf{Peak Temperature.}
Let $T(\mathcal{X}, \mathcal{P})$ be the steady-state thermal distribution under placement $\mathcal{X}$ and power profile $\mathcal{P}$. The thermal objective minimizes the peak temperature
\begin{equation}
    T_p(\mathcal{X}) = \max \big( T(\mathcal{X}, \mathcal{P}) \big).
\end{equation}

\textbf{Interposer Footprint.} Distinct from prior formulations that ignore the footprint, we define a composite footprint objective that captures both the footprint area and the aspect ratio. Specifically, the footprint area is 
\begin{equation}
    A(\mathcal{X}) = W_{bb} \times H_{bb},
\end{equation}
where $W_{bb}$ and $H_{bb}$ denote the bounding box width and height. The aspect ratio is defined as
\begin{equation}
    AR(\mathcal{X}) = \max\!\left(\frac{W_{bb}}{H_{bb}},\; \frac{H_{bb}}{W_{bb}}\right).
\end{equation}
The goal is to minimize footprint area while keeping $AR$ close to a target value $AR_{\text{tgt}}$ (e.g., $AR_{\text{tgt}} = 1$ for a square). The concrete realization of this bi-objective as a unified spatial cost map is detailed in Section~\ref{sec:footprint_mask}.\vspace{-5pt}
\section{The FAPlace Framework}

\begin{figure*}[t]
\centering
\includegraphics[width=2\columnwidth]{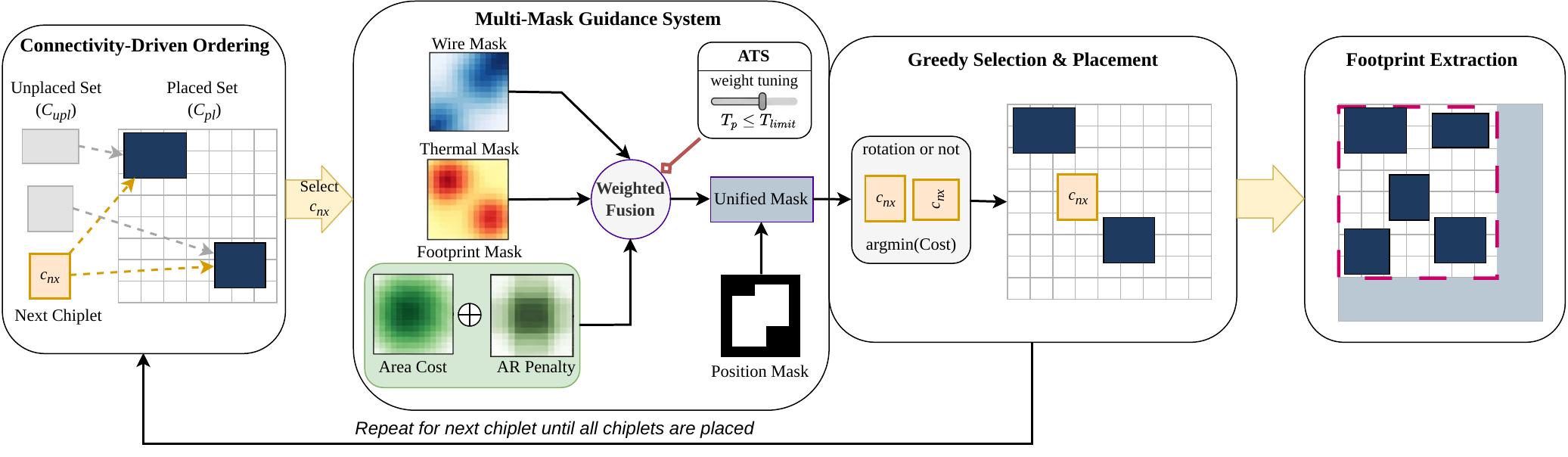}\vspace{-5pt}
\caption{Overview of the FAPlace framework. Chiplets are sequentially placed on a sufficiently large canvas guided by a unified mask synthesized from wirelength, thermal, and the proposed footprint masks.}\vspace{-5pt}
\label{fig:overview}
\end{figure*}

FAPlace is a sequential, mask guided placement framework that determines chiplet locations sequentially on a sufficiently large canvas. As illustrated in Fig.~\ref{fig:overview}, the framework consists of five components: (1) a connectivity-driven ordering strategy, (2) a multi-mask guidance system featuring the novel footprint mask, (3) a unified mask synthesis with greedy selection, (4) footprint extraction, and (5) an adaptive thermal-spacing mechanism to satisfy thermal constraints.\vspace{-5pt}

\subsection{Placement on a Large Canvas}
Unlike conventional methods that confine placement within a pre-specified interposer boundary, FAPlace initializes on a deliberately oversized canvas. The rationale is twofold: first, a large canvas eliminates the infeasibility issues caused by an undersized interposer; second, it decouples the placement quality from the canvas specification, allowing the actual footprint to be determined entirely by the mask-guided optimization. After all chiplets are placed, the layout is shrunk to a tight bounding box and thermally re-evaluated under the compact dimensions. The canvas discretization follows a uniform $G \times G$ grid, and $2G^2$ candidate locations are evaluated at each step to account for both standard and $90^\circ$-rotated orientations.\vspace{-5pt}

\subsection{Connectivity-Driven Ordering}
We employ a connectivity-driven strategy: the chiplet with the largest area is selected as the seed. Subsequently, at each iteration, the unplaced chiplet with the strongest total connection to the already-placed set is selected:
\begin{equation}
    c_{\text{nx}} = \mathop{\arg\max}\limits_{c_i \in \mathcal{C}_{\text{upl}}} \sum_{c_j \in \mathcal{C}_{\text{pl}}} \lambda_{ij},
\end{equation}
where $\lambda_{ij}$ is the connection weight between chiplets $c_i$ and $c_j$, $\mathcal{C}_{\text{pl}}$ and $\mathcal{C}_{\text{upl}}$ are the placed and unplaced sets, respectively. Ties are broken by selecting the candidate with the larger area.\vspace{-5pt}

\subsection{Footprint Mask}\label{sec:footprint_mask}
The footprint mask is the central technical contribution of this work. It encodes a spatial inductive bias that jointly drives chiplets toward compact configurations with a well-proportioned bounding box. For each candidate grid cell $(i,j)$, the mask evaluates the geometric consequence of placing the next chiplet $c_{\text{nx}}$ at that location.

\subsubsection{Area Cost}
Let $A_{\text{curr}}$ denote the bounding box area of the already placed chiplets. When $c_{\text{nx}}$ is centered at $(i,j)$, the new bounding box becomes:
\begin{equation}
\begin{aligned}
W_{\text{new}}(i,j) &= \max(x_{\max}, i+\tfrac{w_{\text{nx}}}{2}) - \min(x_{\min}, i-\tfrac{w_{\text{nx}}}{2}), \\
H_{\text{new}}(i,j) &= \max(y_{\max}, j+\tfrac{h_{\text{nx}}}{2}) - \min(y_{\min}, j-\tfrac{h_{\text{nx}}}{2}),
\end{aligned}
\label{eq:bb}
\end{equation}
where $(x_{\min}, x_{\max}, y_{\min}, y_{\max})$ are the current bounding box extremes. The normalized area expansion cost is:
\begin{equation}
    R_A(i,j) = \frac{W_{\text{new}} \cdot H_{\text{new}} - A_{\text{curr}}}{w_{\text{nx}} \cdot h_{\text{nx}}}.
\label{eq:area_cost}
\end{equation}

\subsubsection{Aspect Ratio Penalty}
The AR penalty quantifies how much the prospective bounding box deviates from the target shape:
\begin{equation}
    R_{AR}(i,j) = \left( \max\!\left(\frac{W_{\text{new}}}{H_{\text{new}}},\; \frac{H_{\text{new}}}{W_{\text{new}}}\right) - AR_{\text{tgt}} \right)^{\!2}.
\label{eq:ar_cost}
\end{equation}
The quadratic form imposes a mild penalty near the target and an increasingly severe one for extreme deviations, reflecting the practical impact of AR on manufacturability.\vspace{-5pt}

\begin{figure}[htbp]
\centering
\includegraphics[width=0.35\columnwidth]{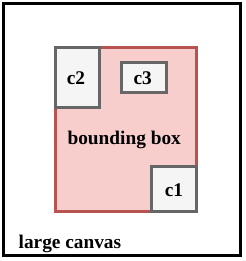}\vspace{-5pt}
\caption{Illustration of the bounding box.} \vspace{-10pt}
\label{fig:shape_mask}
\end{figure}

\subsubsection{Mask Fusion}
Both cost components are independently normalized to $[0,1]$ and fused into the footprint mask:
\begin{equation}
    \mathcal{M}_{F}(i,j) = (1-\eta) \cdot \hat{R}_A(i,j) + \eta \cdot \hat{R}_{AR}(i,j),
\label{eq:shape_mask}
\end{equation}
where $\hat{\cdot}$ denotes min-max normalization and spatial bias weight $\eta \in [0,1]$ controls the trade-off between area compactness and AR compliance.


\subsection{Wire Mask and Thermal Mask}
To guide wirelength and thermal optimization, FAPlace employs two additional masks, inspired by Maskplace~\cite{lai2022maskplace}.\vspace{-5pt}

\subsubsection{Wire Mask}
For the next chiplet $c_{\text{nx}}$, the Wire Mask approximates the routing cost at each candidate location based on the Manhattan distance to connected, already placed chiplets:
\begin{equation}
    \mathcal{M}_{WL}(i,j) = \sum_{c_k \in \mathcal{C}_{\text{pl}}} \lambda(c_{\text{nx}}, c_k) \cdot \left( |i - x_k| + |j - y_k| \right),
\end{equation}
where $(x_k, y_k)$ are the coordinates of placed chiplet $c_k$ and $\lambda(\cdot,\cdot)$ denotes the connection weight.\vspace{-5pt}

\subsubsection{Thermal Mask}
For high-power chiplets (those exceeding a power threshold $p_{\text{th}}$), the Thermal Mask leverages the steady-state temperature distribution $T(\mathcal{X}, \mathcal{P})$ computed by HotSpot~\cite{zhang2015hotspot} for the currently placed chiplets:
\begin{equation}
\mathcal{M}_T(i,j) =
\begin{cases}
T(i,j), & \text{if } p(c_{\text{nx}}) > p_{\text{th}} \\
0, & \text{otherwise}
\end{cases}
\end{equation}
This selectively activates thermal guidance only when placing high power chiplets, avoiding unnecessary constraints on low-power ones.\vspace{-5pt}

\subsection{Unified Mask and Greedy Selection}
The individual masks are normalized and synthesized into a Unified Mask:
\begin{equation}
    \mathcal{M}_{\text{uni}} = (1-\beta)\cdot\alpha \cdot \hat{\mathcal{M}}_{WL} + \beta \cdot \hat{\mathcal{M}}_{T} + (1-\beta)\cdot\gamma \cdot \hat{\mathcal{M}}_{F},
\label{eq:unified}
\end{equation}
where $\alpha = \gamma=0.5$.

To ensure physical validity, a Position Mask $\mathcal{M}_{\text{pos}}$ is applied to filter out infeasible locations, including overlaps or boundary violations:
\begin{equation}
\mathcal{M}_{\text{pos}}(i,j) =
\begin{cases}
0, & \text{if } (i,j) \in \Omega \\
+\infty, & \text{otherwise}
\end{cases}
\end{equation}
where $\Omega$ is the set of feasible positions. The greedy selection then identifies the globally optimal position and orientation:
\begin{equation}
    (x_t^*, \theta_t^*) = \mathop{\arg\min}_{x_t,\; \theta \in \{0^\circ, 90^\circ\}} \left( \mathcal{M}_{\text{uni}}(x_t, \theta) + \mathcal{M}_{\text{pos}}(x_t, \theta) \right).
\end{equation}
This deterministic, single pass procedure places all $N$ chiplets in exactly $N$ steps.\vspace{-8pt}

\subsection{Footprint Extraction}\label{sec:tight_canvas}
Once all chiplets have been placed on the sufficiently large canvas, the interposer footprint is extracted and the layout is thermally evaluated under realistic dimensions.

After the last chiplet $c_N$ is placed, the final bounding box dimensions $W_{\text{new}}$ and $H_{\text{new}}$ are obtained following Eq.~\eqref{eq:bb}. The interposer size is then set to
\begin{equation}
    L_{\text{intp}} = \left\lceil \frac{\max(W_{\text{new}},\; H_{\text{new}})}{g} \right\rceil \cdot g,
\end{equation}
where $g$ is the grid granularity and the ceiling operation ensures alignment with the discretization. All chiplet coordinates are shifted to re-center the layout within this compact $L_{\text{intp}} \times L_{\text{intp}}$ region. This step converts the placement on the oversized canvas into a concrete, fabrication-ready interposer specification.\vspace{-5pt}

\subsection{Adaptive Thermal-Spacing}
To guarantee compliance with a peak temperature limit $T_{\text{limit}}$, FAPlace employs an adaptive thermal-spacing (ATS) mechanism. ATS treats the thermal weight $\beta$ as a tunable parameter and wraps around the placement and tight canvas evaluation pipeline: at each iteration, the full sequential placement is executed with the current $\beta$, followed by footprint extraction. A binary search is then performed to identify the minimum $\beta^*$ that satisfies the constraint:
\begin{equation}
    \beta^* = \min \left\{ \beta \in [0, 1] \;\middle|\; T_p(\beta) \le T_{\text{limit}} \right\}.
\end{equation}
A lower $\beta$ prioritizes compact placement at the expense of higher temperature; ATS efficiently identifies the tightest packing that remains thermally safe. 
\section{Experiments}

\subsection{Experimental Setup}
We evaluate FAPlace on a benchmark suite comprising two real-world industry architectures, MultiGPU~\cite{9474011} (Sys~1) and CPU-DRAM~\cite{kannan2015enabling} (Sys~2), along with three additional synthetic systems of increasing complexity, as summarized in Table~\ref{tab:benchmarks}. We compare against TAP-2.5D~\cite{9474011}, a representative SA-based thermal-aware placement method.\vspace{-10pt}
\begin{table}[htbp]
    \centering
    \caption{Benchmark Design Statistics}\vspace{-10pt}
    \label{tab:benchmarks}
    \begin{tabular}{l|cccc}
        \toprule
        \textbf{Design} & \textbf{\# Chiplets} & \textbf{\# Nets} & \textbf{Avg. Deg.} & \textbf{Max Deg.} \\
        \midrule
        Sys 1  & 6 & 6 & 2.00 & 3 \\
        Sys 2  & 8 & 8 & 2.00 & 3 \\
        Sys 3  & 20 & 20 & 2.00 & 3 \\
        Sys 4  & 28 & 34 & 2.43 & 5 \\
        Sys 5  & 36 & 42 & 2.33 & 5 \\
        \bottomrule
    \end{tabular}
\end{table} \vspace{-10pt}

Wirelength is computed using a multi-commodity flow formulation solved via MILP with IBM ILOG CPLEX v12.8. Peak temperature is obtained from HotSpot v6.0~\cite{zhang2015hotspot}. All metrics, including wirelength, peak temperature, footprint area, and AR, are reported on the tight bounding box canvas after footprint extraction.

FAPlace operates on a sufficiently large canvas without any footprint specification, and the interposer footprint emerges as an output. For baseline methods that require a pre-defined canvas, we use the interposer size produced by FAPlace. This represents a favorable setting for the baselines, as they are given a near-optimal canvas derived from FAPlace's own optimization.

The target AR is set to $AR_{\text{tgt}} = 1.0$ (square). The spatial bias weight $\eta = 0.4$ for Sys 1 and $\eta = 0.2$ for the rest of designs.\vspace{-5pt}

\subsection{Main Results}
We exclude other prior methods from comparison as their implementations are not publicly available. Table~\ref{tab:comparison} presents the comparison between TAP-2.5D and FAPlace. Recall that TAP-2.5D is given FAPlace's output footprint as its input canvas, representing a favorable scenario for the baseline. Despite this advantage, FAPlace achieves superior results across all key metrics.
FAPlace reduces wirelength by approximately $2\times$ on average across all benchmarks, with particularly pronounced improvements on Sys~3 and Sys~5. Both methods achieve comparable thermal profiles, with TAP-2.5D averaging only $1.02\times$ higher peak temperature, confirming that FAPlace's compact placements do not sacrifice thermal quality. Meanwhile, FAPlace is approximately $4\times$ faster than TAP-2.5D, as its deterministic single pass mask-guided greedy selection avoids the iterative stochastic search inherent to SA.

\begin{table}[htbp]
    \centering
    \caption{Performance Comparison: TAP-2.5D vs. FAPlace. $RT$ denotes the runtime and $L_{\text{intp}}$ is the side length of the smallest square interposer that encloses the placement. TAP-2.5D is given FAPlace's output $L_{\text{intp}}$ as its input canvas.}\vspace{-10pt}
    \label{tab:comparison}
    \begin{tabular}{lc|ccc|ccc}
        \toprule
        \multirow{3}{*}{Design} & \multirow{3}{*}{\makecell{$L_{\text{intp}}$\\(mm)}} & \multicolumn{3}{c|}{TAP-2.5D} & \multicolumn{3}{c}{FAPlace} \\
        \cmidrule(lr){3-5} \cmidrule(lr){6-8}
         & & \makecell{$RT$\\(h)} & \makecell{$T_p$\\($^\circ\mathrm{C}$)} & \makecell{$WL$\\(m)} & \makecell{$RT$\\(h)} & \makecell{$T_p$\\($^\circ\mathrm{C}$)} & \makecell{$WL$\\(m)} \\
        \midrule
        Sys 1 & 56 & 1.1 & 90.6 & 122 & 0.2 & 90.6 & 95 \\
        Sys 2 & 43 & 1.4 & 98.7 & 242 & 0.2 & 95.9 & 130 \\
        Sys 3 & 64 & 2.3 & 89.1 & 276 & 0.3 & 88.3 & 79 \\
        Sys 4 & 39 & 3.1 & 73.9 & 40 & 1.2 & 73.8 & 39 \\
        Sys 5 & 34 & 3.5 & 72.9 & 43 & 1.1 & 70.2 & 19 \\
        \midrule
        \multicolumn{2}{c|}{Avg.} & 3.99$\times$ & 1.02$\times$ & 1.99$\times$ & 1$\times$ & 1$\times$ & 1$\times$ \\
        \bottomrule
    \end{tabular}
\end{table}

\begin{figure}[htbp]
\centering
\begin{subfigure}{0.24\columnwidth}
  \includegraphics[width=\linewidth]{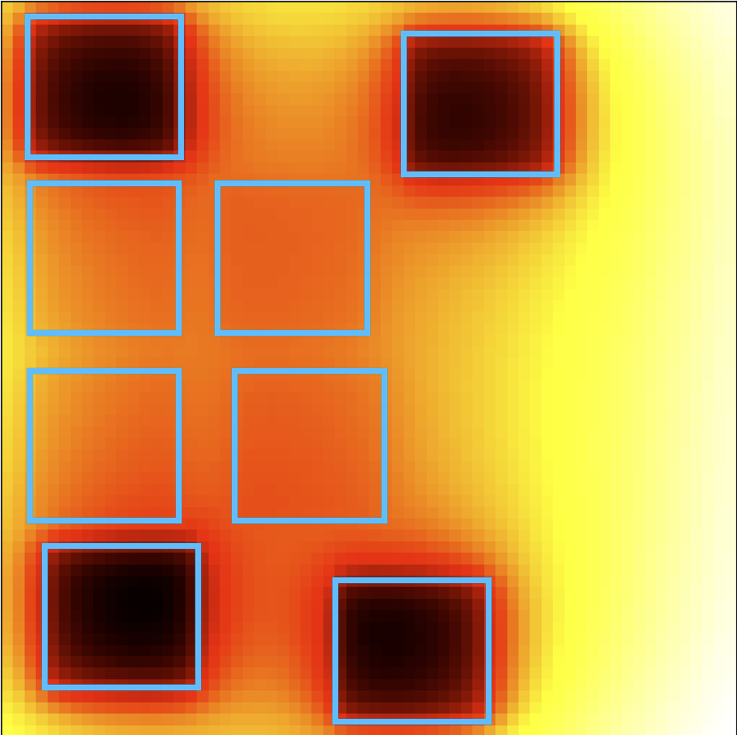}
  \caption{\footnotesize{TAP-2.5D, Sys 2}}
\end{subfigure}\hfill
\begin{subfigure}{0.24\columnwidth}
  \includegraphics[width=\linewidth]{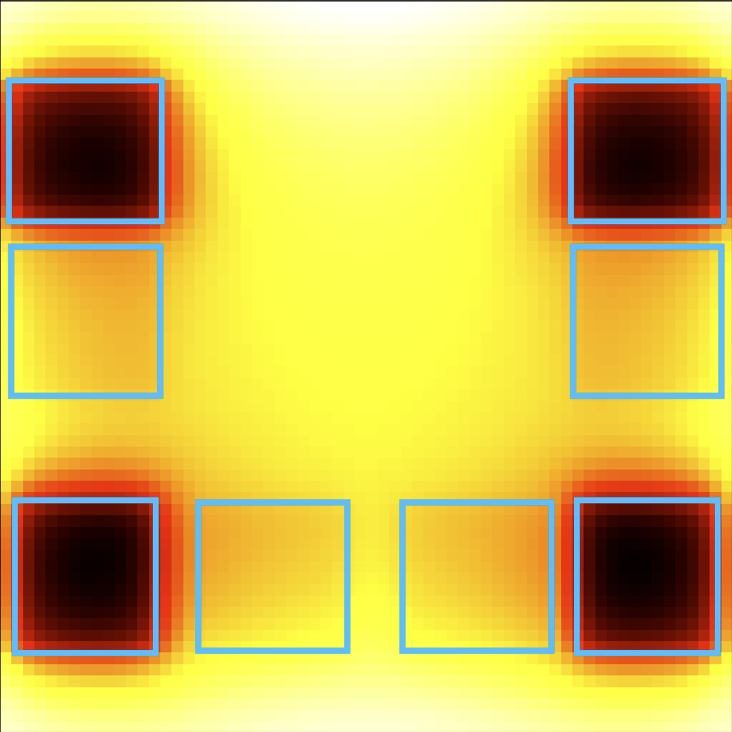}
  \caption{\footnotesize{FAPlace, Sys 2}}
\end{subfigure}\hfill
\begin{subfigure}{0.24\columnwidth}
  \includegraphics[width=\linewidth]{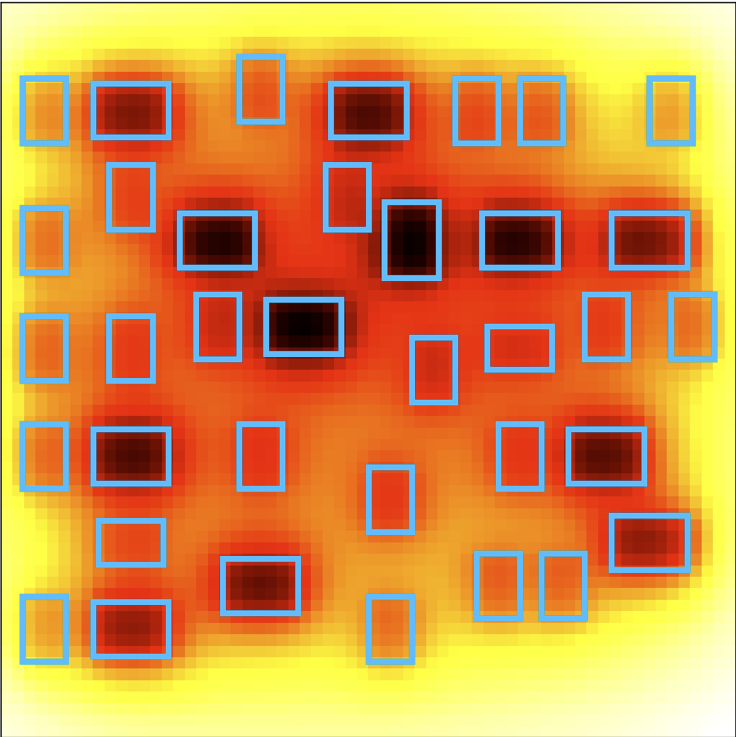}
  \caption{\footnotesize{TAP-2.5D, Sys 5}}
\end{subfigure}\hfill
\begin{subfigure}{0.24\columnwidth}
  \includegraphics[width=\linewidth]{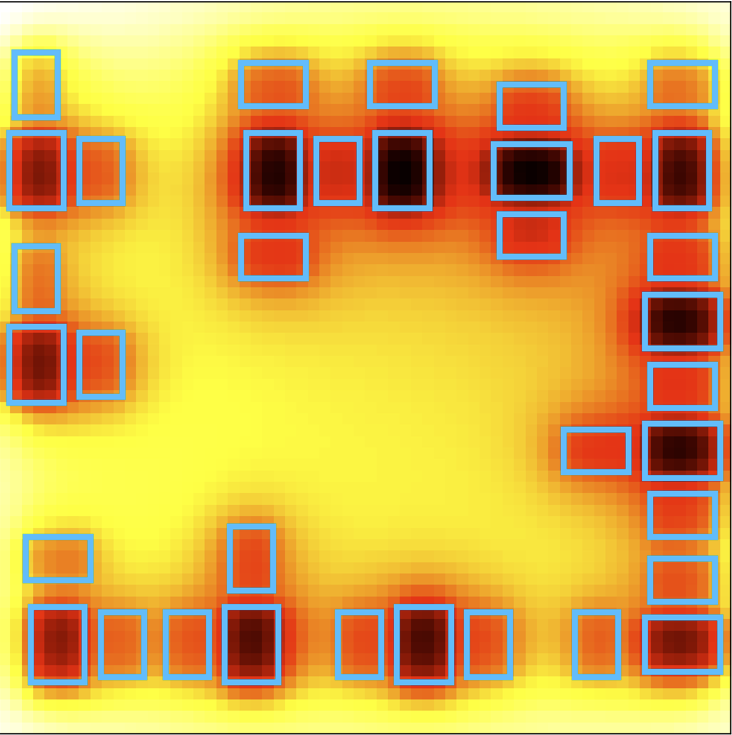}
  \caption{\footnotesize{FAPlace, Sys 5}}
\end{subfigure}\vspace{-10pt}
\caption{Thermal maps with chiplet placement overlays for different methods.} \vspace{-10pt}
\label{fig:canvas}
\end{figure}

Figure~\ref{fig:canvas} visualizes the thermal layouts for Sys~2 and Sys~5. While TAP-2.5D suffers from localized heating due to clustered or uniformly scattered placements, FAPlace strategically separates heat sources. FAPlace effectively mitigates peak temperatures and facilitates heat dissipation across the interposer.

\subsection{Evaluation on a Large Canvas}
The main comparison in Table~\ref{tab:comparison} provides TAP-2.5D with FAPlace's optimized $L_{\text{intp}}$ as its input canvas, effectively supplying a geometric prior that removes the need for TAP-2.5D to determine the interposer footprint on its own. To eliminate this advantage and establish a level playing field, we run TAP-2.5D on the same sufficiently large canvas used by FAPlace, using Sys~1 as a case study. For a fair comparison, we select a TAP-2.5D placement whose peak temperature closely matches that of FAPlace, ensuring that differences in wirelength, bounding box area, and aspect ratio are not artifacts of differing thermal budgets.

As shown in Table~\ref{tab:large_canvas}, without the geometric prior, TAP-2.5D exhibits significant degradation across all metrics. These results demonstrate that TAP-2.5D relies heavily on a well-specified interposer footprint to produce compact, well-proportioned placements. In contrast, FAPlace, guided by the footprint mask, autonomously converges to a tight and near-square layout without requiring any pre-defined boundary specification.

\begin{table}[htbp]
    \centering
    \caption{Performance comparison on a sufficiently large canvas for Sys~1.}\vspace{-10pt}
    \label{tab:large_canvas}
    \begin{tabular}{l|cccc}
        \toprule
        \makecell{Method} & \makecell{$T_p$\\($^\circ\mathrm{C}$)} & \makecell{$WL$\\(m)} & \makecell{$A$\\(cm$^2$)} & $AR$ \\
        \midrule
        TAP-2.5D & 90.3 & 339 & 106 & 1.40 \\
        FAPlace  & 90.6 & 95 & 29 & 1.12 \\
        \bottomrule
    \end{tabular}
\end{table}\vspace{-10pt}

\subsection{Ablation Study}
Table~\ref{tab:ablation} shows the effect of the footprint mask. Without it, the placement can still achieve comparable wirelength and peak temperature, yet the overall layout quality degrades significantly. Specifically, the average aspect ratio increases by 1.80$\times$, and the interposer side length grows by 1.40$\times$. This demonstrates that the footprint mask serves as an effective shape constraint without compromising electrical or thermal performance.

The necessity of both mask components is further highlighted by specific design cases. On Sys~1, removing the mask actually yields a smaller bounding box area, yet the aspect ratio surges from 1.12 to 3.07. This elongation forces the layout onto an 85 mm interposer instead of a 56 mm one. Such an elongated shape cannot be efficiently accommodated on a square interposer, entirely negating the apparent area savings. Conversely, on Sys~2, although the aspect ratio remains unaffected without the mask, the bounding box area nearly doubles from 15 to 28 cm$^2$, expanding the required interposer from 43 mm to 58 mm. These cases illustrate that both the area cost and the aspect ratio penalty are essential for the footprint mask to enforce a compact and manufacturable outline.
\begin{table}[htbp]
    \centering
    \caption{Ablation Study: Effect of the Footprint Mask. $A$ denotes the bounding box area of the placement, and $L_{\text{intp}}$ is the side length of the smallest square interposer that encloses it.}\vspace{-10pt}
    \label{tab:ablation}
    \resizebox{\linewidth}{!}{
    \begin{tabular}{l|ccccc|ccccc}
        \toprule
        \multirow{3}{*}{Design} & \multicolumn{5}{c|}{w/o Footprint Mask} & \multicolumn{5}{c}{FAPlace} \\
        \cmidrule(lr){2-6} \cmidrule(lr){7-11}
         & \makecell{$T_p$\\($^\circ\mathrm{C}$)} & \makecell{$WL$\\(m)} & \makecell{$A$\\(cm$^2$)} & \makecell{$L_{\text{intp}}$\\(mm)} & $AR$ & \makecell{$T_p$\\($^\circ\mathrm{C}$)} & \makecell{$WL$\\(m)} & \makecell{$A$\\(cm$^2$)} & \makecell{$L_{\text{intp}}$\\(mm)} & $AR$ \\
        \midrule
        Sys 1 & 90.5 & 91 & 24 & 85 & 3.07 & 90.6 & 95 & 29 & 56 & 1.12 \\
        Sys 2 & 96.3 & 142 & 28 & 58 & 1.23 & 95.9 & 130 & 15 & 43 & 1.23 \\
        Sys 3 & 88.3 & 73 & 54 & 98 & 1.80 & 88.3 & 79 & 36 & 64 & 1.12 \\
        Sys 4 & 73.7 & 38 & 14 & 48 & 1.69 & 73.8 & 39 & 11 & 39 & 1.16 \\
        Sys 5 & 70.5 & 19 & 9 & 47 & 2.47 & 70.2 & 19 & 10 & 34 & 1.13 \\
        \midrule
        Avg. & 1.00$\times$ & 0.99$\times$ & 1.27$\times$ & 1.40$\times$ & 1.80$\times$ & 1$\times$ & 1$\times$ & 1$\times$ & 1$\times$ & 1$\times$ \\
        \bottomrule
    \end{tabular}
    }
\end{table}\vspace{-10pt}

\subsection{Impact of Area Cost and AR Penalty}
\begin{figure}[htbp]
\centering
\includegraphics[width=0.6\columnwidth]{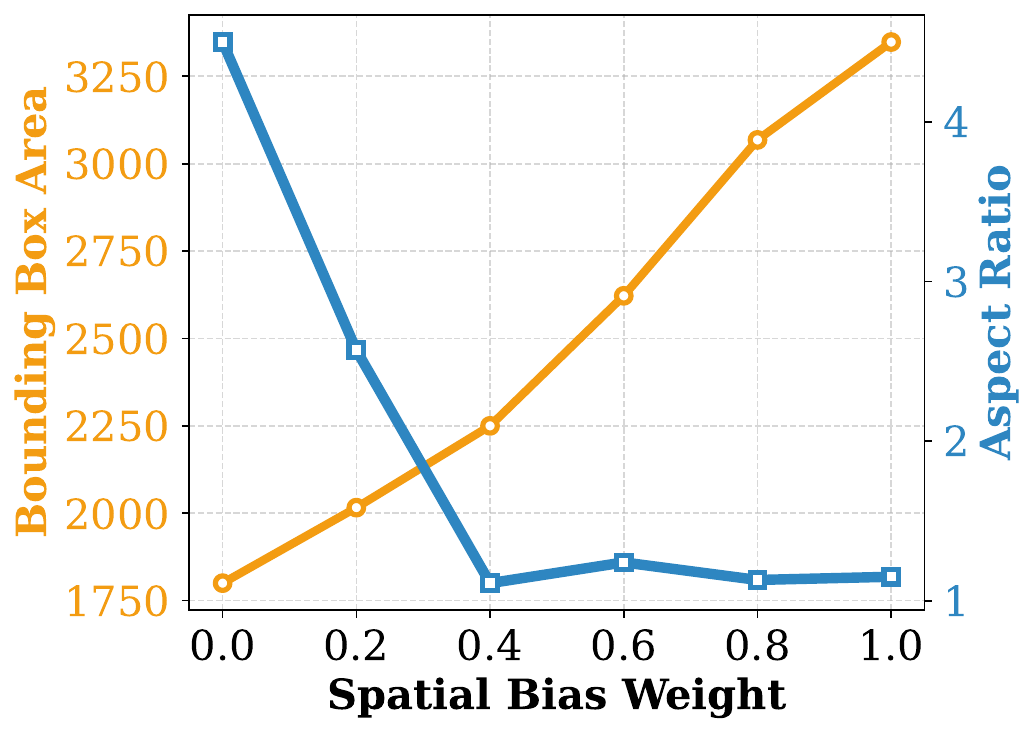}\vspace{-10pt}
\caption{Quantitative trade-off analysis between area cost and aspect ratio penalty.} 
\label{fig:ar_area}
\end{figure}

\begin{figure}[htbp]
\centering
\begin{tabular}{ccc}
\includegraphics[width=0.26\columnwidth]{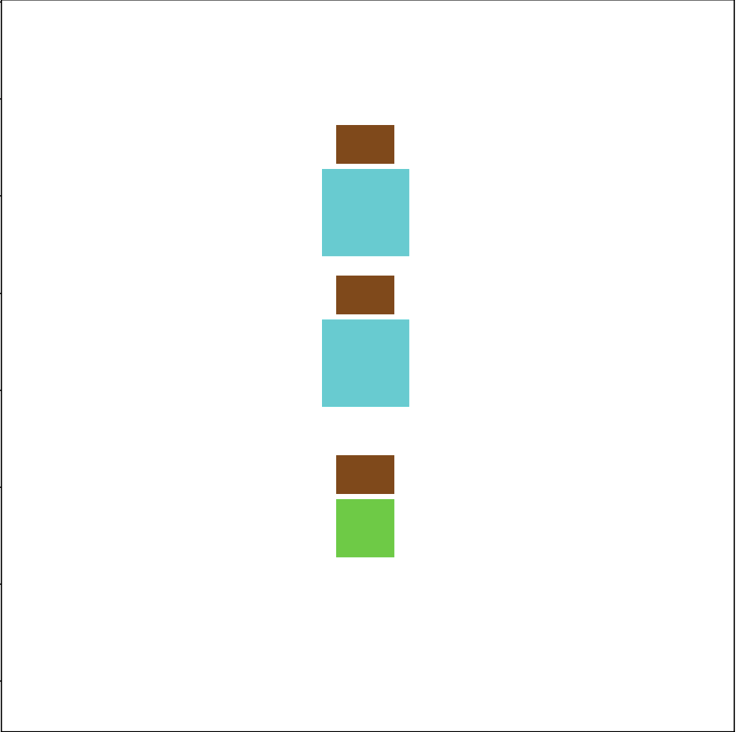} &
\includegraphics[width=0.26\columnwidth]{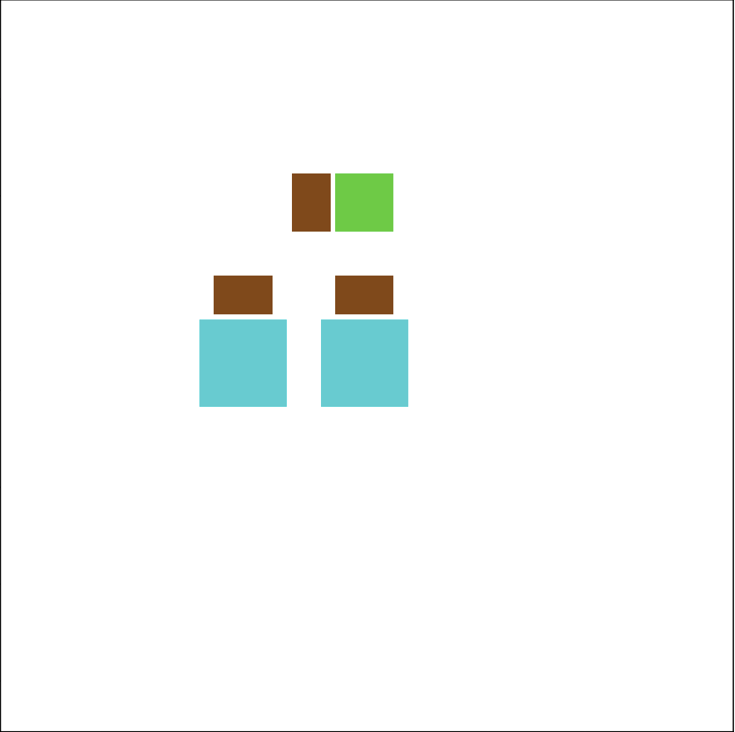}  &
\includegraphics[width=0.26\columnwidth]{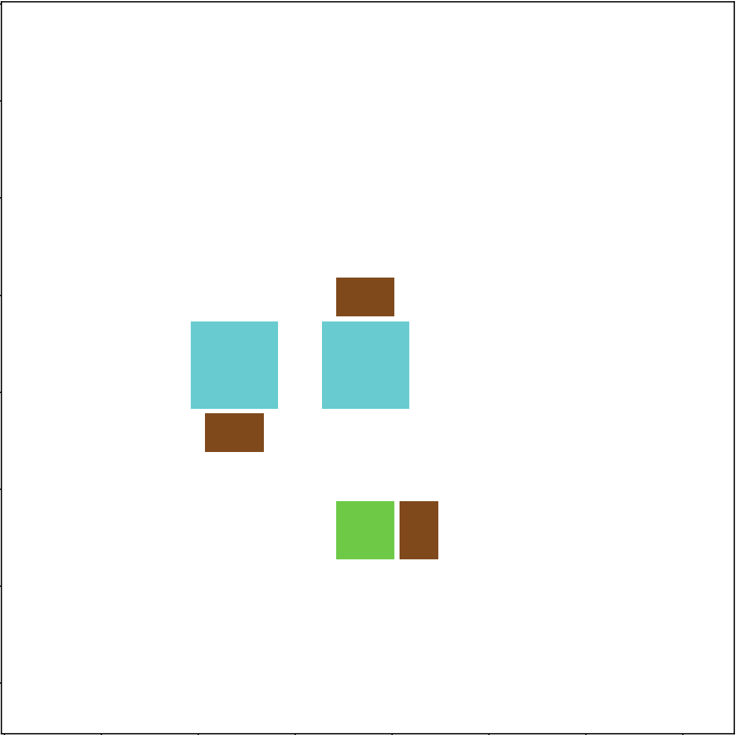} \\
(a) $\eta{=}0$ & (b) $\eta{=}0.4$ & (c) $\eta{=}0.8$ 
\end{tabular}\vspace{-10pt}
\caption{Placement layouts under varying spatial bias weights for Sys 1 on a large canvas.}\vspace{-10pt}
\label{fig:canvas_AR}
\end{figure}

We conduct a sensitivity analysis to examine the individual contributions of the area cost and AR penalty within the footprint mask. Specifically, we vary the weighting coefficient $\eta$ and report the resulting bounding box area and aspect ratio on Sys 1 in Figure~\ref{fig:ar_area}. Figure~\ref{fig:canvas_AR} shows the corresponding layout visualizations.
When the AR penalty is entirely disabled ($\eta = 0$), the placement is driven predominantly by bounding box area optimization. Although this configuration achieves the absolute minimum bounding box area, it results in an extreme aspect ratio of approximately 4.5. Visually, this manifests as an elongated, quasi-one-dimensional chain of chiplets, which is highly impractical for real-world manufacturing.
As $\eta$ increases, the AR penalty effectively regularizes the layout geometry. We observe a sharp decline in the AR, which converges to the target value of approximately 1.0 (i.e., a square footprint) once $\eta \ge 0.4$.

\subsection{Impact of Adaptive Thermal-Spacing}
\begin{figure}[htbp]
\centering
\includegraphics[width=0.6\columnwidth]{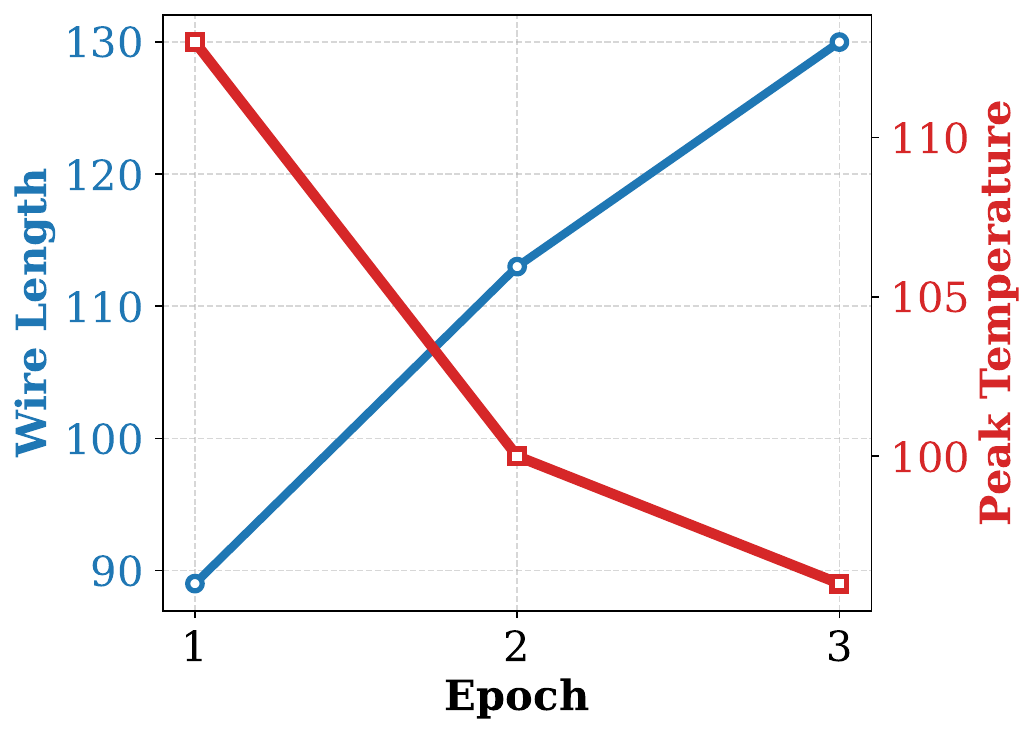}\vspace{-10pt}
\caption{Convergence of wirelength and peak temperature under the ATS mechanism on Sys 2.}\vspace{-10pt}
\label{fig:ATS}
\end{figure}

\begin{figure}[htbp]
\centering
\begin{tabular}{ccc}
\includegraphics[width=0.26\columnwidth]{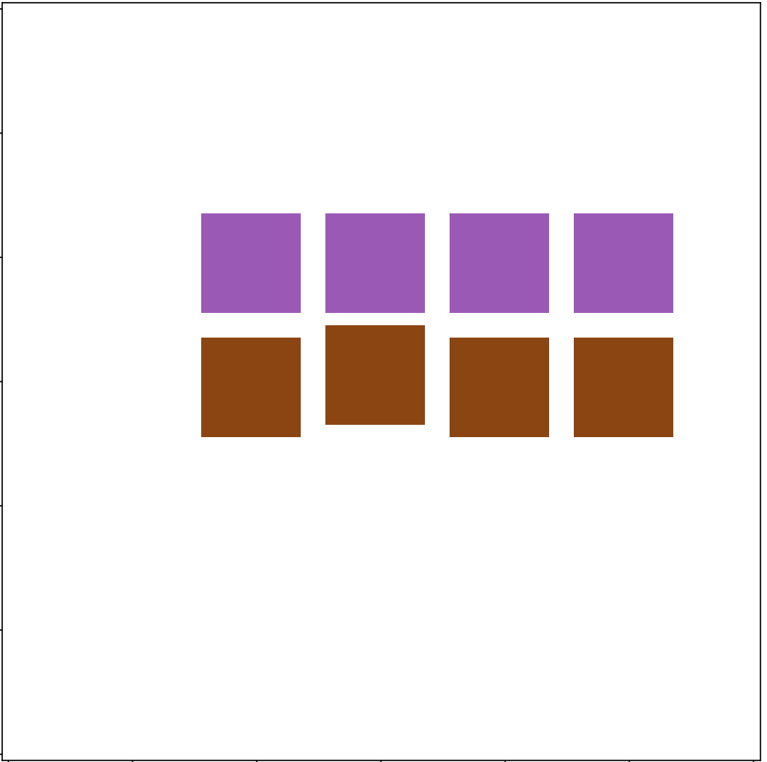} &
\includegraphics[width=0.26\columnwidth]{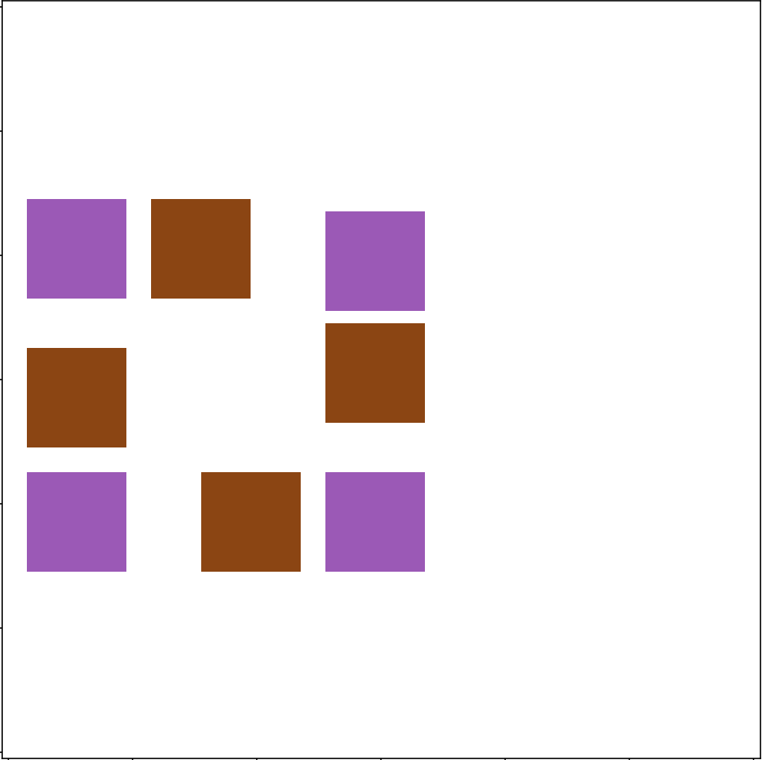} &
\includegraphics[width=0.26\columnwidth]{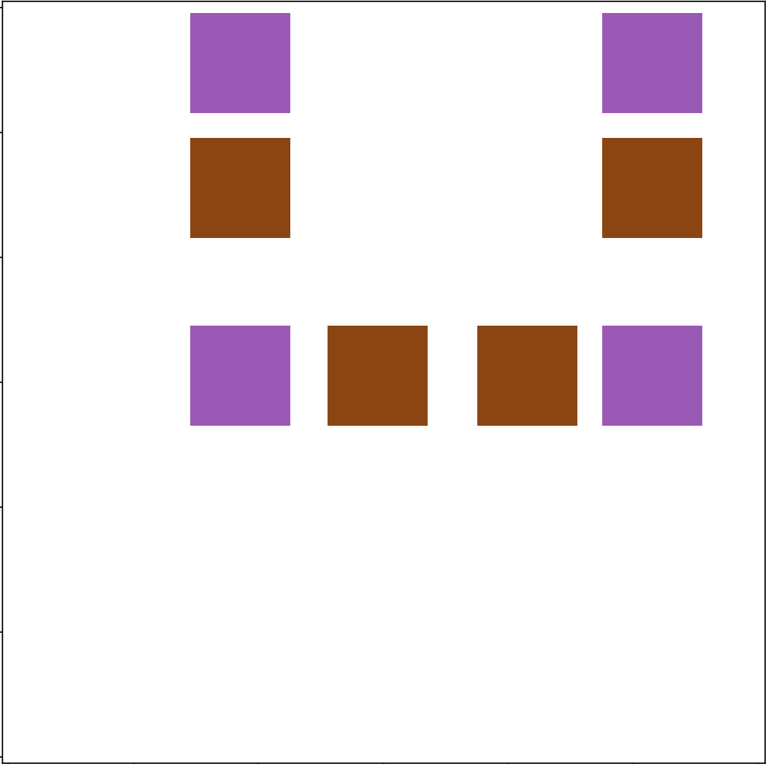} \\
 Epoch 1 & Epoch 2 & Epoch 3 
\end{tabular}\vspace{-10pt}
\caption{Evolution of placement layouts driven by the ATS mechanism for Sys~2 on a large canvas.}\vspace{-10pt}
\label{fig:canvas_sys2}
\end{figure}

To validate the responsiveness of the ATS mechanism, we tracked both the performance in Figure~\ref{fig:ATS} and the corresponding spatial layout in Figure~\ref{fig:canvas_sys2} during the placement of Sys~2, targeting a peak temperature limit of 96$^\circ\mathrm{C}$.
Driven by the ATS, the layout dynamically evolves to resolve thermal violations. Initially, a low thermal weight forces a dense core, which minimizes wirelength but causes a severe thermal violation. As ATS dynamically increases thermal weight to penalize localized heating, the layout expands, dropping the temperature to 100$^\circ\mathrm{C}$ at the expense of increased wirelength. By Epoch 3, ATS converges on the optimal weight. The layout transforms into a dispersed, U-shaped perimeter configuration, successfully satisfying the rigorous 96$^\circ\mathrm{C}$ limit while establishing the minimal acceptable wirelength for this specific thermal constraint.

\vspace{-5pt}

\section{Conclusion}
In this paper, we presented FAPlace, a footprint-aware mask-guided sequential placement framework that eliminates the circular dependency between chiplet placement and interposer footprint specification in 2.5D systems. By operating on a sufficiently large canvas and introducing a novel footprint mask that fuses area compactness with an aspect ratio penalty, FAPlace allows the optimal interposer footprint to emerge naturally as an output of the optimization. Integrated with wirelength and thermal guidance masks, the framework achieves holistic multi-physics optimization through a deterministic, single pass process. Experimental results across diverse benchmarks demonstrate that FAPlace reduces wirelength and footprint area while achieving near-unity aspect ratios, without compromising thermal performance.\vspace{-5pt}

\begin{acks}
This research is supported by the Agency for Science, Technology and Research (A*STAR) under its MTC Programmatic Funds (Grant No. M23M3b0064). 
\end{acks}

\bibliographystyle{ACM-Reference-Format}
\bibliography{references}

\end{document}